\documentclass{article}
\usepackage{amsfonts,amssymb,amsmath,amsthm}
\usepackage{times}		     
\title{Semidefinite programming and arithmetic circuit evaluation}
\date{September, 30, 2005}
\author{Sergey P. Tarasov\and Mikhail N. Vyalyi}

\let\leq\leqslant			    		 
\let\geq\geqslant			    		 
\let\eps\varepsilon
\let\al\alpha
\renewcommand*\P{\ensuremath{\mathrm {P}}}
\newcommand*\NP{\ensuremath{\mathrm {NP}}}
\newcommand*\coNP{\ensuremath{\mathrm {coNP}}}
\newcommand*\BPP{\ensuremath{\mathrm {BPP}}}
\newcommand*\EXPTIME{\ensuremath{\mathrm {EXPTIME}}}

\newcommand*{\poly}{\mathop{\mathrm{poly}}}
\newcommand*\PSPACE{\ensuremath{\mathrm {PSPACE}}}
\newcommand*\NC{\ensuremath{\mathrm {NC}}}

\def\lcm{\mathop{\mathrm{lcm}}\nolimits}
\def\poly{\mathop{\mathrm{poly}}\nolimits}
\def\tail{\mathop{\mathrm{tail}}\nolimits}

\newtheorem{thm}{Theorem}
\newtheorem{lemma}{Lemma}
\newtheorem{prop}{Proposition}
\newtheorem{st}{Statement}
\def\QQ{\ensuremath{\mathbb Q}}
\def\ZZ{\ensuremath{\mathbb Z}}
\def\RR{\ensuremath{\mathbb R}}

\def\D{\ensuremath{\mathcal D}}
\def\cP{\ensuremath{\mathcal P}}
\def\N{\ensuremath{\mathcal N}}
\def\L{\ensuremath{\mathcal L}}
\def\R{\ensuremath{\mathcal R}}
\def\A{\ensuremath{\mathcal A}}
\def\B{\ensuremath{\mathcal B}}
\def\Sc{\ensuremath{\mathcal S}}
\def\AC{\ensuremath{\mathrm {AC}}}

\begin{document}
\maketitle

\begin{abstract} 
A rational number can be naturally presented by an \emph{arithmetic
computation}  (AC): a sequence of elementary arithmetic operations
starting from a fixed constant, say $1$.  The asymptotic complexity
issues of such a representation are studied e.g. in \cite{BCSS,
Koiran} in the framework of the algebraic complexity theory 
over arbitrary
field.

Here we study a related problem of the complexity of performing
arithmetic operations and computing elementary predicates, e.g. ``=''
or ``$\geq$'', on rational numbers given by AC.

In the first place, we prove that AC can be 
efficiently simulated by the
exact semidefinite programming (SDP).

Secondly, we give a \BPP-algorithm for the equality predicate.

Thirdly, we put $\geq$-predicate into the complexity class 
\PSPACE.

We conjecture that $\geq$-predicate is hard to compute. This
conjecture, if true, would clarify the complexity status of the exact
SDP --- a well known open problem in the field of mathematical
programming.

\end{abstract}

Algorithmic complexity provides a general framework to analyze
complexity of computational problems. It works for many cases and
gives results that are important for practical applications.
Nevertheless, some basic assumptions of the theory are strange from
practical point of view. We are sure that 
a linear time algorithm 
running in time
$10^{1000}n$ cannot be realized in our Universe. Also, 
some widely used algorithms have
exponential running time in the worst case. 

Numerical algorithms are especially in striking disagreement with
complexity theory. Probably, the most popular exponential algorithm is
the simplex algorithm for linear programming.
It is widely used despite the existence of polynomial algorithms that were
found after pioneering breakthrough of L.Khachiyan (see, e.g.
\cite{KhLP, Kar, Re}).

The complexity analysis of the semidefinite programming problem
involves even more difficulties. SDP is often considered as tractable
due to various approximate algorithms. SDP is a convex optimization
problem so the ellipsoid method can be applied to solve it
approximately as well as a variety of interior points
methods~\cite{GLS,NN,Re98,WSV}.  But there are amazingly few results
on complexity of the SDP problem.

Khachiyan and Porkolab~\cite{PKh} found polynomial time algorithm for
the SDP problem when the dimension is  fixed. They established
doubly exponential bounds on the  solutions for the general
SDP problem and on the discrepancies of infeasible programs. 

For our further considerations the most interesting is Ramana's
result~\cite{Ra97}. He developed the exact duality theory for
SDP. Ramana's dual program can be constructed in polynomial time. His
analogue of Farkas lemma has an
immediate complexity-theoretic corollary:
a complement to the SDP feasibility problem (SDFP)
can be reduced to SDFP itself. It implies that SDFP
cannot be \NP-complete unless $\NP=\coNP$. More generally, SDFP should
belong to a complexity class that is closed under 
complement. Examples 
are \P, $\NP\cap\coNP$, \BPP,  \PSPACE.
(For definitions of these classes and other useful information on
complexity theory see, e.g., the Sipser's book~\cite{Sipser}.)

Here we address the exact SDFP problem. 
The problem is to check that intersection of the cone of positive
semidefinite
matrices with some affine subspace of matrices is not empty. The
subspace involved is defined by generators that are matrices with
rational entries. It is well known that some pathological examples
exist for the SDFP. For instance, it is possible that feasible program
has only doubly exponential solutions and that an infeasible program
could have doubly exponentially small discrepancy. These examples show
that in some cases a polynomially bounded machine can not simply write
a solution to SDFP. Of course, this does not prevent checking the
existence of a solution in polynomial time.

Motivated by these examples, we introduce a new problem:
comparison of arithmetic computations.  It seems to be
interesting in it's own right and concerns a nonstandard way of
representing integers and rationals.  The common way to represent
integers is to use a positional system.  Among positional systems the
binary system is the simplest and the most natural. Binary
representation of a number $N$ is a string of length $\theta(\log
N)$. This bound is optimal due to a counting argument.  But it is also
possible to encode numbers in such a way that some numbers are encoded
by very short strings. In this case we speak of a \emph{succinct
representation} of an integer.
A natural way 
for succinct representation of a rational number $r$ 
is to 
use  an
\emph{arithmetic computation} or an \emph{ arithmetic circuit} (AC). By
definition, AC is a sequence of the elementary arithmetic operations
starting from a fixed constant, say $1$, 
and generating
$r$ as output.

To our knowledge, the complexity issues of the AC representation were
studied primarily in the asymptotic setting in the framework of the
algebraic complexity theory over a field (see, e.g.  \cite{SS,
Koiran}).  Namely, let $\tau(k)$ be the minimum number of arithmetic
operations required to build the integer $k$ from the constant, say
$1$, i.e. the length of the minimal AC computing $k$. A sequence $x_k$
of integers is said to be ``ultimately easy to compute'' if there
exists another sequence $a_k$ and polynomial $p(\cdot)$ such that
$\tau(a_kx_k)\leq p(\log k)$ for all $k$ (for instance, in can be
shown that the sequence $2^k$ is ultimately easy to compute).
Otherwise the sequence is said to be ``ultimately hard to compute''.
Counting argument shows that ultimately hard to compute sequences do
exist. A central open problem is to show that some explicit sequences,
say, $n!$, $\lfloor (3/2)^n \rfloor$ are ultimately hard to
compute. For instance, if $n!$ is ultimately hard to compute then $\P_C
= \NP_C$ over the field of complex numbers~\cite{SS}.

It is worth to mention that the complexity of AC is related to
a circuit model of computation for
polynomials that was introduced and studied by Valiant~\cite{Val}. 
The Valiant's model turns to AC  by substitution
constants in polynomials.  Some interesting applications
of this observation are discussed in~\cite{Koiran}.

In this paper we 
explicitly treat 
a related problem of the complexity
of performing arithmetic operations and computing elementary
predicates, e.g. ``='' or ``$\geq$'', on rational numbers represented
by AC.  An immediate inspection shows that if numbers are repesented
by AC then to perform an elementary arithmetic operation one should
simply merge the AC of the operands in the appropriate way, thus
arithmetical operations are easy to perform, despite the evident fact
that 
AC representation is succinct 
for some numbers
(e.g. $2^{2^n}$ can be obtained by repeated squaring). On the
other hand, --- it is unclear how to 
efficiently
compute elementary
predicates ``='' or ``$\geq$'', 
i.e.
how to check 
efficiently 
whether two AC compute the same value or how to compute
the maximal value.
In the sequel we denote these predicates by $\AC_=$ and $\AC_\geq$
respectively.

Our main contributions are: we prove that
some restricted version of 
AC can be 
efficiently simulated by the
exact semidefinite programming 
(SDP) and construct a polynomial reduction of $\AC_\geq$ to SDFP;
we give a
\BPP-algorithm for the equality predicate; we put $\geq$-predicate into
the complexity class \PSPACE.

Unfortunately, we are unable at the moment to prove any lower bounds
for the $\geq$-predicate. Any nontrivial result in this direction
would imply
lower bounds for the exact SDP --- one of 
the main
open problems in the field of mathematical programming.

The rest of paper is organized as follows. In Section~\ref{CAC} we
give a definition of the circuit representation for rationals and state the
basic results concerning it. 
Also, this Section
contains technical results 
about a choice of a basis for AC.  In Section~\ref{red}
we construct a reduction of $\AC_\geq$ to SDFP.  Sections~\ref{t1},
\ref{t2} contain proofs of $\AC_=\in \BPP$ and $\AC_\geq\in\PSPACE$.
In the last Section~\ref{open} we discuss open
questions around the $\AC_\geq$ and SDFP problems.

\section{AC representation of numbers}
\label{CAC}

Let $\B$ be a finite collection of functions of type $\QQ^k\to \QQ$ ($k$ may
vary). It is called a \emph{basis\/}.
A \emph{circuit} over basis~$\B$ is a sequence
$\Sc=s_0,s_1,\dots,s_\ell$ of assignments such that $s_0:=1$ and for
each $i\geq 1$ there exist $j,k<i$ such that $s_i:=s_j*s_k$ and $*\in \B$.
The \emph{size} $\ell(\Sc)$ of  circuit $\Sc$ is the number~$\ell$.  

The most natural basis consists of four arithmetic operations 
$\{+,-,/,\cdot\}$.  Circuits over this basis are called
\emph{arithmetic circuits\/}. Circuits over the basis
$\{+,-,\cdot\}$ are called \emph{division-free}
circuits. \emph{Monotone} (arithmetic) circuits are circuits over the
basis $\{+,\cdot\}$.

We will represent rationals by circuits. For each circuit we define the
\emph{value} of the circuit by induction. We will use notation
$v(\Sc)$ for the value of circuit $\Sc$.
Value of 1 is 1. The value of
a circuit $\Sc=s_0,s_1,\dots,s_\ell$ where $s_\ell=s_j*s_k$ is 
$v(\Sc_j)*v(\Sc_k)$. Here $\Sc_j=s_0,s_1,\dots,s_j$.
Note that each prefix of a circuit is a circuit by definition. If some
operations can not be performed (e.g. a division by~0) the value of
such a circuit is undefined.
Let $X=v(\Sc)$. We say that such  AC \emph{represents} $X$.  

It is easy to see that arithmetic circuit representation can be much more
compressed than the usual binary representation. 
A circuit
$\Sc=1,s_1,s_2,\dots, s_\ell$ where $s_1=1+1$ and $s_{j+1}=s_{j}\cdot
s_j$ for $j\geq1$ represents $2^{2^{\ell-1}}$. On the other hand this
example is asymptotically optimal.  The following statement can be
easily verified by induction.

\begin{st}
  If $p/q=v(\Sc)$ where $p,q$ are integers then  $\max \{p,q\}=O(2^{2^\ell})$.
\end{st}

\subsection{The complexity of equality and inequality predicates over AC}

Implementation of arithmetic operations with rationals represented by 
arithmetic circuits is straightforward and can be done
in linear time. Formally we write
\begin{equation*}
  \begin{split}
  \Sc(X*Y)=\Sc(X), \tail(\Sc(Y)), s_{\ell(\Sc(X))+\ell(\Sc(Y)-1)},\\
  \text{where }s_{\ell(\Sc(X))+\ell(\Sc(Y))-1}:=
  s_{\ell(\Sc(X))}*s_{\ell(\Sc(X))+\ell(\Sc(Y))-2}.
  \end{split}
\end{equation*}
Here $\tail(\Sc)$ means a circuit $\Sc$ without the starting~1.

But how difficult may be the computation of the equality predicate and
of the inequality predicate?  We state these algorithmic problems
formally. 
We always assume that a circuit is represented
by a list of triples $(*,j,k)$ where $*\in \B$ and $j,k$ are positive
integers. The $n$th element in the list corresponds to an assignment
$s_n=s_j*s_k$.  Thus a circuit of size $\ell$ is written as a
$O(\ell\log\ell)$ binary word.

\medskip

{\boldmath\textbf{Predicate $\AC_=(\B)$.}} It is true for a pair of
circuits $\Sc_1$, $\Sc_2$ over the basis $\B$ iff $v(\Sc_1)=v(\Sc_2)$.

{\boldmath\textbf{Predicate $\AC_\geq(\B)$.}} 
It is true for a pair of
circuits $\Sc_1$, $\Sc_2$ over the basis $\B$ iff $v(\Sc_1)\geq v(\Sc_2)$.

\medskip

\textbf{Note.}  If $\B$ contains division then these predicates are
partially defined (value of some circuits can be undefined). Partially
defined predicates are called \emph{promise problems\/}.  Most
complexity classes can be easily redefined to include promise problems and
most results remain true in this, more general, setting.  We omit a
discussion of promise problems but indicate that their use is safe in our
considerations.

We denote the equality and the inequality predicates 
over the arithmetic basis 
by $\AC_=$ (resp. $\AC_\geq$).

It is clear that both predicates fall into the complexity class
\EXPTIME. In fact, we are able to place them into lower levels of
computational hierarchy.

 \begin{thm}\label{th=}
  $\AC_=\in\BPP$.
 \end{thm}

In other words, the equality check can be performed by probabilistic
Turing machine in polynomial time. 

We are unable to give an exact characterization of the computational complexity
of the second predicate. Computing $\AC_\geq$ looks as a
computationally hard problem. 
An obvious way to solve it is to make all calculations indicated in
the circuits that form input of the problem. Using binary
representation we need exponentially large memory to do it. Using
modular arithmetic it is possible to solve $\AC_\geq$ in polynomial
memory.

\begin{thm}\label{upper}
  $\AC_\geq\in \PSPACE$.
\end{thm}

The proof of Theorem~\ref{upper} uses some constructions from
an $\NC^1$ algorithm
for comparison integers in modular arithmetic~\cite{NC1, DL91}.

\subsection{Equivalent bases}\label{bases}

Two bases $\B_1$ ¨ $\B_2$ are called \emph{``$=$'' equivalent\/}
(respectively, \emph{``$\geq$'' equivalent\/}) iff the predicates
$\AC_=(\B_1)$ and $\AC_=(\B_2)$ (respectively, $\AC_\geq(\B_1)$ and
$\AC_\geq(\B_2)$) are mutually polynomially reducible.

\begin{thm}\label{eq-bases}
  The following bases are ``$=$'' and ``$\geq$'' equivalent:
  arithmetic, division-free, monotone and $\{+,x\mapsto x^2/2\}$.
\end{thm}

The last basis in the list is added for technical purposes. It is used
in the reduction of the problem $\AC_\geq$ to the feasibility problem
for semidefinite programming.

Reductions of all mentioned bases to the arithmetic basis are
straightforward. To prove Theorem~\ref{eq-bases} we establish
reductions in the opposite direction.

Note that if a basis contains a subtraction then general predicate
$\AC_\geq(\B)$ is reducible to its particular case when one of the compared
values is zero. Indeed, suppose we are going to compare $v(\Sc_1)$
and $v(\Sc_2)$. We can merge the circuits $\Sc_1$, $\Sc_2$   into one
circuit. This merged circuit contains all assignments of $\Sc_1$,
$\Sc_2$  and ends by the assignment $d:=a-b$, where $a$ and $b$ are
the last assignments in the circuits $\Sc_1$, $\Sc_2$.
The same argument can be also applied to the predicate $\AC_=(\B)$.

All reductions described below have similar form. A circuit $\Sc$ over some
basis is converted to a circuit $\Sc'$ over another basis using step-by-step
substitution of constant-sized groups of assignments instead of each 
assignment in $\Sc$.

\begin{lemma}\label{Lemma1}
   $\AC_\geq(\{+,-,/,\cdot\})$
  (resp. $\AC_=(\{+,-,/,\cdot\})$)
 is reducible to $\AC_\geq(\{+,-,\cdot\})$
 (resp. $\AC_\geq(\{+,-,\cdot\})$).
\end{lemma}

\begin{proof}
Informally speaking, the lemma is very simple: we can keep and transform numerators
and denominators separately. Below we present a more detailed description
of the reduction.

Let $\Sc$ be a circuit of size $\ell$ over the arithmetic basis. We
construct a circuit $\Sc'$ of size $O(\ell)$ over division-free
basis in the following way.

The circuit $\Sc'$ consists of four sequences of assignments $\A^1$, $\A^2$,
$\N$, $\D$. 
For the assignment  $s_i:=s_j\pm s_k$ in  $\Sc$
add the assignments
\begin{equation*}
  \A^1_i:=\N_j\cdot \D_k,\ 
  \A^2_i:= \D_j\cdot \N_k,\ 
  \D_i:=\D_j\cdot\D_k,\ 
  \N_{i}:=\A^1_i\pm \A^2_i.
\end{equation*}
Similarly, 
for the assignment  $s_i:=s_j\cdot s_k$ in  $\Sc$
add the assignments
\begin{equation*}
  \D_i:=\D_j\cdot\D_k,\ 
  \N_{i}:=\N_j\cdot\N_k,
\end{equation*}
and for the assignment  $s_i:=s_j/ s_k$ in $\Sc$ 
add the assignments
\begin{equation*}
  \D_i:=\D_j\cdot\N_k,\ 
  \N_{i}:=\N_j\cdot\D_k.
\end{equation*}

The last assignment in the circuit $\Sc'$ is
\begin{equation*}
  s'_{N}:=\N_\ell\cdot\D_\ell.
\end{equation*}

It is easy to see that 
\begin{equation*}
  v(\Sc)=\frac{v(\N_\ell)}{v(\D_\ell)}. 
\end{equation*}
So, $v(\Sc)\geq 0$ iff $s'_N\geq 0$.

Note that due to our assumptions the case 
$\D_\ell=0$ is impossible. So, the same reduction is valid for the
predicate $\AC_=$.
\end{proof}

\begin{lemma}
    $\AC_\geq(\{+,-,\cdot\})$ (resp. $\AC_=(\{+,-,\cdot\})$)
  is reducible to 
  $\AC_\geq(\{+,\cdot\})$ (resp. $\AC_\geq(\{+,\cdot\})$). 
\end{lemma}

\begin{proof} Informally, we do the same trick 
  as above
  using equalities
\begin{gather}
  (A-B)+(C-D)=(A+C)-(B+D),\label{+}\\
  (A-B)-(C-D)=(A+D)-(B+C),\label{-}\\
  (A-B)\cdot(C-D)=(A\cdot C+B\cdot D)-(B\cdot C+A\cdot D).\label{x}
\end{gather}

Now let consider the details of
the reduction.

At first we convert an input $(\Sc_1,\Sc_2)$ of
$\AC_\geq(\{+,-,/,\cdot\})$ into $(\Sc,0)$ as it was explained above.

Then we construct two circuits $\L$, $\R$ such that
$v(\Sc)=v(\L)-v(\R)$. So, $v(\Sc)\geq0$ iff
$v(\L)\geq v(\R)$ as well as $v(\Sc)=0$ iff
$v(\L)= v(\R)$. Again, the same reduction will work for both
predicates. 
The size of  $\L$, $\R$ will be $O(s(\Sc))$ and
they will be the circuits over  $\{+,\cdot\}$. 

Both circuits $\L$ and $\R$ consist of six sequences of assignments
$\L^1$, $\L^2$, $\L^3$, $\R^1$, $\R^2$, $\R^3$. 

For the assignment  $s_i:=s_j+s_k$ 
in  $\Sc$ add assignments to $\R$
\begin{equation*}
  \L^1_i:=\L^1_j+ \L^1_k,\ 
  \R^1_i:=\R^1_j+ \R^1_k
\end{equation*}
(see Eq.~\eqref{+}).

For the assignment $s_i:=s_j - s_k$ in $\Sc$
add the assignments to $\R$
\begin{equation*}
  \L^1_i:=\L^1_j+ \R^1_k,\ 
  \R^1_i:=\R^1_j+ \L^1_k
\end{equation*}
(see Eq.~\eqref{-}).

For the assignment  $s_i:=s_j \cdot s_k$ in  $\Sc$
add the assignments to $\R$ 
\begin{equation*}
  \begin{split}
  &\L^3_i:=\L^1_j\cdot \L^1_k,\ 
  \L^2_i:=\R^1_j\cdot \R^1_k,\ 
  \L^1_i:=\L^2_i+\L^3_i,\\
  &\R^3_i:=\R^1_j\cdot \L^1_k,\ 
  \R^2_i:=\L^1_j\cdot \R^1_k,\ 
  \R^1_i:=\R^2_i+\R^3_i.
  \end{split}
\end{equation*}
(see Eq.~\eqref{x}).

The circuit $\L$ has the same structure except $\L_i$ and
$\R_i$ assignments (or groups of assignments) are
interchanged. By induction, we see that for each $i$
the value of $s_i$ is $v(\L^1_i)-v(\R^1_i)$.
\end{proof}

\begin{lemma}
    $\AC_\geq(\{+,\cdot\})$ 
  (resp.  $\AC_=(\{+,\cdot\})$)
  is reducible to $\AC_\geq(\{+,x\mapsto x^2/2\})$
  (resp. $\AC_=(\{+,x\mapsto x^2/2\})$).  
\end{lemma}
\begin{proof}
Let $\Sc$ be a circuit over the monotone basis. We
construct a circuit $\Sc'$  over the basis $\{+,x\mapsto x^2/2\}$
consisting of 10 series of assignments
$\cP$, $\N$, $\A^t$, $t\in[1,8]$ such that for each $k$ the equality
\begin{equation}\label{diff}
  v(s_k)=v(\cP_k)-v(N_k)
\end{equation}
holds. The construction is  step-by-step substitution as in the above
lemmas.

For the assignment $s_i:=s_j+ s_k$ in  $\Sc$ add the assignments
\begin{equation}\label{add}
  \cP_i:=\cP_j+ \cP_k,\ 
  \N_{i}:=\N_j+\N_k
\end{equation}
and for the assignment $s_i:=s_j\cdot s_k$ add the assignments
\begin{equation}\label{ass}
  \begin{split}
    &A^1_i:=\cP_j+\cP_k,\ A^2_i:= \N_j+\N_k,\
    A^3_i:=\cP_j+\N_k,\ A^4_i:=\N_j+\cP_k,\\
    &A^5_i:=(A^1_i)^2/2,\ A^6_i:=(A^2_i)^2/2,\ 
    A^7_i:=(A^3_i)^2/2,\ A^8_i:=(A^4_i)^2/2,\\
    &\cP_i=A^5_i+A^6_i,\\
    &\N_i=A^7_i+A^8_i.
  \end{split}
\end{equation}
In the latter case the following  equations hold
\begin{equation}\label{sum0squares}
  \begin{split}
    &v(\cP_i)=(v(\cP_j)+v(\cP_k))^2/2+(v(\N_j)+v(\N_k))^2/2,\\
    &v(\N_i)=(v(\cP_j)+v(\N_k))^2/2+(v(\N_j)+v(\cP_k))^2/2.
  \end{split}
\end{equation}

Eq.~\eqref{diff} is verified
by induction using Eq.~\eqref{sum0squares}
\begin{multline}
  v(s_i)=(v(\cP_j)-v(\N_j))(v(\cP_k)-v(\N_k))=\\
  (v(\cP_j)\cdot v(\cP_k)+v(\N_j)\cdot v(\N_k))-\\
  (v(\cP_j)\cdot v(\N_k)+v(\N_j)\cdot v(\cP_k))=\\
  ((v(\cP_j)+v(\cP_k))^2/2+(v(\N_j)+v(\N_k))^2/2)-\\
  ((v(\cP_j)+v(\N_k))^2/2+(v(\N_j)+v(\cP_k))^2/2)=\\
  v(\cP_i)-v(\N_i).
\end{multline}

Now we are able to construct a reduction. Take an instance $(\Sc_1,\Sc_2)$ of
the problem $\AC_\geq(\{+,\cdot\})$. 
Convert  the circuits $\Sc_1$, $\Sc_2$ into the circuits $\Sc^{1}$,
$\Sc^2$ over the basis $\{+,x\mapsto x^2/2\}$ as described above. Join
them into one circuit $\Sc$. The circuit $\Sc'$ is an extension of $\Sc$
by  the  assignment $f':=\cP_{s_1}^1+\N_{s_2}^2$. Similarly, the
circuit $\Sc''$ is an extension of $\Sc$ by the assignment 
$f'':=\cP_{s_2}^2+\N_{s_1}^1$.  Here $s_1$ is a size of $\Sc_1$ and
$s_2$ is a size of $\Sc_2$. 

A reduction is given by the mapping
$
(\Sc_1,\Sc_2)\mapsto (\Sc',\Sc'')
$.
From Eq.~\eqref{diff} we see that it also works for both types of
predicates mentioned in the lemma.
\end{proof}

\section{Reduction of the problem\boldmath{} $\AC_\geq$ to SDFP}\label{red}

Linear optimization on the intersection of the cone of positive 
semidefinite matrices with an affine  subspace of matrices is called 
\emph{semidefinite programming}  (SDP). Let denote SDFP the corresponding
feasibility problem: to check whether the cone of positive semidefinite
matrices has nonempty intersection with affine subspace of subspace.

Semidefinite feasibility problem (SDFP) can be stated in the following
form.

\textbf{Input:} a list $Q_0,\dots, Q_m$ of symmetric $(n\times n)$
matrices with rational entries. Matrices are represented by lists of
entries, each entry is represented by a pair (numerator, denominator),
integers are given in binary.

\textbf{Output:} `yes' if there exist reals $x_1,\dots,x_m$ such that 
$Q=Q_0+\sum_{i=1}^mx_iQ_i$ is a positive semidefinite matrix, otherwise
the output is `no'.

The proof  below uses Ramana's results on
the exact duality theory for SDP~\cite{Ra97}. 
Ramana found a special form of dual program for SDP. It is called the
extended Lagrange -- Slater dual program (ELSD). Ramana proved that
\begin{itemize}
\item ELSD can be
  constructed from the primal program in polynomial time;
\item if both the primal and the dual are feasible then their optimum
  values are equal.
\end{itemize}

\begin{thm}\label{reduction}
  The problem $\AC_\geq(\{+,x\mapsto x^2/2\})$ is reducible to \textup{SDFP}.
\end{thm}

Applying Theorem~\ref{eq-bases} we have reductions of the problem
$\AC_\geq$ over all bases discussed in the previous section to SDFP.

\begin{proof}
The first step is to represent
a circuit value as an optimal value of some semidefinite program in the 
form
\begin{equation}\label{prog}
  \begin{split}
    &t\to\inf,\\
    &\text{semidefinte conditions on $t$ and other variables.}
  \end{split}
\end{equation}

Let $\Sc$ be a circuit over the basis $\{+,x\mapsto x^2/2\}$. 
We construct a semidefinite program $P(\Sc)$ such that
for each assignment $s_i$ there is a variable $x_i$ in $P(\Sc)$ and a
matrix $M_i$.
Matrices $M_i$ are built as follows
\begin{equation*}
  \begin{aligned}
    &s_i:=s_j+s_k&&\longrightarrow&&
    M_i=\begin{pmatrix}
    x_i-x_j-x_k
    \end{pmatrix}\\
    &s_i:=s_j^2/2&&\longrightarrow&&
    M_i=\begin{pmatrix}
    2x_i&x_j\\ x_j&1
    \end{pmatrix}
  \end{aligned}
\end{equation*}
Note that in the former case $M_i\succeq 0$ iff $x_i\geq x_j+x_k$ and
in the latter $M_i\succeq 0$ iff $x_i\geq x_j^2/2$.

Let $\ell$ be a size of $\Sc$. Consider a SDP
\begin{equation}\label{scheme-prog}
  \begin{split}
    &x_\ell\to\inf,\\
    &M_i\succeq0 \text{ for each }i,\\
    &x_0=1.
  \end{split}
\end{equation}
To convert the program~\eqref{scheme-prog} to the standard form we
replace an equality $x_0=1$ by  inequality
\begin{equation}\label{eq2pos}
  \begin{pmatrix}
    x_0-1&0\\ 0&-x_0+1
  \end{pmatrix}\succeq0. 
\end{equation}

The optimal value of $P(\Sc)$ is the value of $\Sc$.  Indeed, all
operations in the process of circuit evaluation are monotone with
respect to each variable. So, any feasible solution
of~\eqref{scheme-prog} satisfies the condition $x_\ell\geq v(\Sc)$. On
the other hand, $x_i=v(s_i)$ is a feasible solution.

Take now an instance of $\AC_\geq(\{+,x\mapsto x^2/2\})$. It's input is a
pair of circuits $\Sc^{(1)}$, $\Sc^{(2)}$.  The positive answer in the
problem  $\AC_\geq(\{+,x\mapsto x^2/2\})$ means that 
$v(\Sc^{(1)})\geq v(\Sc^{(2)})$ which is equivalent to
$-v(\Sc^{(2)})\geq-v(\Sc^{(1)})$. The value
 $-v(\Sc^{(2)})$ is an optimal value for SDP
\begin{equation}\label{S^2}
  \begin{split}
    -x_\ell^{(2)}&\to\sup,\\
    M_i^{(2)}&\succeq0,\\
    x_0^{(2)}&=1.
  \end{split}
\end{equation}

To represent  $-v(\Sc^{(1)})$ as the infimum of SDP we use the
extended Lagrange--Slater dual program 
to the program of type~\eqref{scheme-prog}. 
In Ramana's paper ELSD  has a mixed form (linear equations are
permitted). To convert it to the standard form each linear equation 
should be
replaced by a  positive semidefinite condition on a $(2\times 2)$
matrix as  in Eq.~\eqref{eq2pos}. 
After convertion  ELSD can be written as follows
\begin{equation}\label{S^1}
  \begin{split}
    c(Y)&\to\inf,\\
    D&\succeq0.
  \end{split}
\end{equation}
Here $c(Y)$ is a linear functional on dual variables $Y$.

Thus, the program
\begin{equation}\label{prog-fin}
  \begin{split}
    -x_\ell^{(2)}-c(Y)&\geq0,\\
    M_i^{(2)}&\succeq0,\\
    x_0^{(2)}&=1,\\
    D&\succeq0.
  \end{split}
\end{equation}
is feasible iff $-v(\Sc^{(2)})\geq-v(\Sc^{(1)})$. 

This gives a reduction of 
$\AC_\geq(\{+,x\mapsto x^2/2\})$ to SDFP.
\end{proof}

\section{The proof of Theorem \ref{th=}}\label{t1}

The idea is to check the equality modulo random number. 

It is easy to compute a remainder of a circuit value modulo `short'
number (represented in binary). Addition, multiplication and integer
division are made in polynomial time by standard algorithms.

We show that if two circuits have different
values then with big enough probability  
they have different residues modulo random number. 

We present a \BPP-algorithm for the monotone basis. By
theorem~\ref{eq-bases} we conclude that $\AC_=\in\BPP$.

\medskip

\textbf{The probabilistic algorithm for\boldmath{} $\AC_=(\{+,\cdot\})$.}

\begin{description}
\item [Input:] a pair  $(\Sc_1,\Sc_2)$ where $\Sc_1$, $\Sc_2$ 
  are circuits over the monotone basis. Let $\ell$ be the maximal size
  of the circuits.
\item [Step 1.] Set $B$ to $2^{2\ell}$. 
\item [Step 2.] Choose a random integer $m$ from the uniform distribution on
  the interval $[1,B]$.
\item [Step 3.] Compute $r_1=v(\Sc_1)\bmod m$, $r_2=v(\Sc_2)\bmod m$
  by making all operations indicated in the circuits $\Sc_1$, $\Sc_2$
  modulo~$M$.
\item [Step 4.] If $r_1=r_2$ then output `$v(\Sc_1)=v(\Sc_2)$' else
  output `$v(\Sc_1)\ne v(\Sc_2)$'.
\end{description}

It is clear from the description that
the algorithm uses $2\ell$ random bits and runs in polynomial time. 

\textbf{Claim 1.} If $v(\Sc_1)=v(\Sc_2)$ then the algorithm outputs 
`$v(\Sc_1)\ne v(\Sc_2)$' with probability~0.

This is clear.

\textbf{Claim 2.} If $v(\Sc_1)\ne v(\Sc_2)$ then the algorithm outputs
`$v(\Sc_1)\ne v(\Sc_2)$' with probability at least $(2\ell)^{-1}$. 

Claim 2 is derived from Lemma~\ref{lcm} stated below.

To fit the common definition of the class $\BPP$ we need to amplify
success probability by the standard procedure of multiple repetitions
of the algorithm. Claims  show that the gap to be amplified is
$\Omega(n^{-1})$ where $n$ is an input size. So, an
amplification can be made by $\poly(n)$ repetitions and the resulting
algorithm runs in polynomial time. 

Now we need a lower bound of the least common
multiple of integers taken from an interval $[1,B]$ provided that we take 
sufficiently many integers.

\begin{lemma}\label{lcm}
  Let 
  $$
  N_\eps(B)=\min\{\lcm(x_1,\dots,x_r):
  x_i\in[1,B], r>(1-\eps) B\}.$$ 
  Then for all $\eps<(2\ln B)^{-1}$   we have
  \[
  N_\eps(B)>2^{\Omega(B/\ln B)}.
  \]
\end{lemma}

\begin{proof}
  Consider a set of integers $R=\{x_1,\dots,x_r\}\subseteq[1,B]$ such
  that $|R|>(1-\eps)B$. 
  Factorize the least common multiple of these integers 
  $$X=\lcm(x_1,\dots,x_r)=p_1^{a_1}\dots p_s^{a_s}.$$ We are going to
  show that $s=\Omega(B/\ln B)$. Then the lemma will follow from
  trivial bound $X>2^s$.

  The prime number theorem gives an asymptotic 
  \begin{equation}\label{PNT}
    \pi(n)\sim\frac n{\ln n}
  \end{equation}
  for the prime counting function $\pi(n)=\#\{p: p\leq n, p\text{
    prime}\}$. 

  Let $\bar R=[1,B]\setminus\{x_1,\dots,x_r\}$. At least $\pi(B)-s$
  primes in the interval  $[1,B]$ belong to the set $\bar R$. From an
  inequality  $\pi(B)-s\leq B-r$ we conclude that 
  \[
   (1-\eps) B< r\leq B-\pi(B)+s
  \]
  and $s=\Omega(B/\ln B)$ for $\eps<(2\ln B)^{-1}$.
\end{proof}

To complete the proof of Theorem~\ref{th=}
we derive Claim~2 from Lemma~\ref{lcm}.

  W.l.o.g. assume that $v(\Sc_1)> v(\Sc_2)$. From upper bounds of circuit value
  we have
  \begin{equation}\label{Delta}
  \Delta=v(\Sc_1)- v(\Sc_2)\leq 2^{1+2^\ell}.
  \end{equation}
  Let $R$ be a set of integers such that $m\in R$ iff $m\in [1,B]$ and
  $$
  (v(\Sc_1)- v(\Sc_2)) \bmod m=0. 
  $$ 
  Note that if $m\notin R$ then the algorithm outputs `$v(\Sc_1\ne
  v(\Sc_2)$'.

  The set $R$ cannot be large. At first we note that $\Delta$ is a
  multiple of $\lcm_{x\in R} x$. Suppose that
  $\#R>(1-\eps)B$ for $\eps=(2\ell)^{-1}$. We have
  $\eps=(2\ell)^{-1}=(2\log_2B)^{-1}<(2\ln B)^{-1}$.
  Thus, Lemma~\ref{lcm} implies 
  $$
  \Delta\geq \lcm_{x\in R} x
  >2^{\Omega(B/\ln B)}=2^{\Omega(2^{2\ell}/\ell)}>\Omega(2^{2^{1.5\ell}})
  $$
  and we come to the contradiction with~\eqref{Delta}.
  This contradiction shows that $\#R<(1-(2\ell)^{-1})B$. Claim~2
  follows from this bound.

\section{The proof of Theorem \ref{upper}}\label{t2}

We prove Theorem~\ref{upper} by adjusting an $\NC^1$-algorithm for
integer comparison by Davida and Litow~\cite{DL91}. So, we partially 
reproduce arguments from~\cite{NC1,DL91}. 

We start by some notation.
Let  $p_1=3, p_2,\dots, p_m$ be the first $m$ odd primes.
We denote the least nonnegative residue of~$x$ modulo~$n$ by $[x]_n$: 
$x\equiv [x]_n\pmod n$, $0\leq[x]_n<n$. 
For the rest of the section we set
$M=p_1p_2\dots p_m$, $M_i=M/p_i$, $x_i=[x]_{p_i}$,
$\xi_i=[xM_i^{-1}]_{p_i}$.

We denote the fractional part $x-\lfloor
x\rfloor$ of $x$ by  $\{x\}$.

By the Chinese remainder theorem any integer $0\leq x<M$ is uniquely
represented by $x_i$. For any integer $x$ the following equality holds
\begin{equation}\label{Chinese}
  \sum_{i=1}^m M_i\cdot\xi_i =\rho(x)\cdot M+[x]_M.
\end{equation}
Since $M_i\xi_i<M$, we have
$0\leq \rho(x)<m$. The integer~$\rho(x)$ is called the \emph{rank} of
$x$ with respect to $\{p_1,\dots, p_m\}$. 

Dividing both sides of Eq.~\eqref{Chinese}  by~$M$ we get 
\begin{equation}\label{rho1}
  \sum_{i=1}^m\frac{\xi_i}{p_i} =\rho(x)+\frac{[x]_M}{M}.
\end{equation}
Eq.~\eqref{rho1} gives a way to compute rank $\rho(x)$ by
approximating left-hand side.
Let choose an integer $h$ such that  $m/2^h<1/4$. 
For each $i$ we take a $(2^{-h})$-approximation of
$\xi_i/p_i$, i.e. 
\begin{equation}\label{approx}
  \frac{s_i}{2^h}\leq\frac{\xi_i}{p_i}< \frac{s_i+1}{2^h},\
  s_i,h\in\ZZ_+,\quad
  \text{where}\ s_i=\left\lfloor\frac{2^h\xi_i}{p_i}\right\rfloor.
\end{equation}
Let $\sigma(x)=\sum_{i=1}^ms_i2^{-h}$. Then, by summation of
Eqs.~\eqref{approx} for all $i$ we get
\begin{equation}\label{fracpart}
  \{\sigma(x)\}=\rho(x)-  \lfloor \sigma(x)\rfloor + \frac{[x]_M}M-\al
\quad\text{and }0\leq\al<1/4.
\end{equation}

Now we reproduce Lemma 2.3 from~\cite{NC1}.

\begin{prop}\label{facts}
  The following holds.
  \begin{enumerate}
  \item If $\{\sigma(x)\}\leq 3/4$ then $\rho(x)= \lfloor
    \sigma(x)\rfloor $.
  \item If $1/4\leq [x]_M/M\leq 3/4$ then $\{\sigma(x)\}\leq 3/4$ and
    $\rho(x)= \lfloor \sigma(x)\rfloor $.
  \item If $[x]_M/M>1/2$ then $\rho(x)= \lfloor \sigma(x)\rfloor$.
  \end{enumerate}
\end{prop}

All these facts are easily derived from Eq.~\eqref{fracpart}.
Using them we obtain an analogue of Lemma 2.5 from~\cite{NC1}.

\begin{lemma}\label{criterion}
  Let $k^*(x)=\min(k: k\geq0\text{ and }\{\sigma([2^kx]_M)\}\leq
  3/4)$. For any $x$ the following holds.
  \begin{itemize}
  \item   $k^*(x)<\infty$.
  \item $k^*(x)=2^{\poly(m)}$.
  \item Suppose that $k^*(x)>0$. Then $[x]_M/M<1/2$ iff
   $[[2^{k^*(x)}x]_M]_2=0$.
  \end{itemize}
\end{lemma}

\begin{proof}
  For $x$ such that $1/4\leq [x]_M/M\leq 3/4$ we get  $k^*(x)=0$ from
  Proposition~\ref{facts}.2. So, for this case the lemma is trivial.
  
  Suppose that  $[x]_M/M< 1/4$ and $k^*(x)>0$. Choose $\tilde k\geq1$
  such that
  \begin{equation}\label{logbound}
    \frac 1{2^{\tilde k+2}}< \frac{[x]_M}M < \frac 1{2^{\tilde k+1}}\,.
  \end{equation}
  Applying Proposition~\ref{facts}.2 we conclude that
  $\{\sigma([2^{\tilde k}x]_M)\}\leq 3/4$. Therefore $k^*(x)\leq \tilde
  k<\infty$.  The prime number theorem~\eqref{PNT} 
  has an equivalent form
  $$
  p_k\sim k\ln k.
  $$
  It gives a bound $M=2^{\poly(m)}$. Thus $k^*(x)=2^{\poly(m)}$.

  For $k\leq k^*(x)\leq \tilde k$ we use the upper bound
  of~\eqref{logbound} to show that
  $2^k[x]_M<M$. It implies that $[2^{k^*(x)}[x]_M]_M=2^{k^*(x)}[x]_M$
  is even.

  Now suppose that  $ [x]_M/M> 3/4$ and $k^*(x)>0$. 
  Let $y=M-[x]_M=[y]_M$. So, $[y]_M/M<1/4$ and
  $[2^ky]_M=[-2^kx]_M=M-[2^kx]_M$. Repeating the above argument 
  we choose   $\tilde k\geq1$ such that
  \begin{equation}\label{logbound1}
    \frac 1{2^{\tilde k+2}}< \frac{[y]_M}M < \frac 1{2^{\tilde k+1}}\,.
  \end{equation}
  Since $[2^{\tilde k}x]_M=M-[2^{\tilde k}y]_M$, we conclude that
  $k^*(x)=2^{\poly(m)}$ and $[2^{k^*(x)}[x]_M]_M=M-2^{k^*(x)}[y]_M$ is odd.
\end{proof}

The algorithm described below
compares integers represented by circuits over the
division-free basis $\{+,-,\cdot\}$. As was explained above,
in this case comparing values of two circuits is
reduced to comparing a circuit value with~0, i.e. to
computing the sign of the circuit value. 

Let $\Sc$ be a circuit of size~$\ell$. As it was mentioned above, 
  $|v(\Sc)|<2^{2^\ell}$.  
There are $m=\pi(2^{2\ell})-1$ primes from $3$ to $2^{2\ell}$.
Due to the prime number theorem
$$
  2^{1.5\ell}<m<2^{2\ell}.
$$
By the Chinese remainder theorem the residues modulo these primes 
represent uniquely all
circuit values for circuits of size~$\ell$.
Moreover, $|v(\Sc)|<M/4$ in these settings.

\medskip

\textbf{The algorithm to check\boldmath{}  $v(\Sc)>0$.}
  \begin{description}
\item [Input:] $\Sc$~--- a circuit of size $\ell$ over the
  division-free basis.
\item[Step 1.]\label{a1} Let $h=2\ell$.
\item[Step 2.]\label{a3}  Compute $k^*(v(\Sc))$. 
\item[Step 3.]\label{a5} If  $k^*(v(\Sc))>0$ go to Step~6. 
\item[Step 4.]\label{a6} Compute $a_1:=[v(\Sc)]_2$, $a_2:=[[v(\Sc)]_M]_2$.
\item[Step 5.]\label{a7} If $a_1=a_2$ then output `yes' else output `no'.
\item[Step 6.]\label{a-rho} Compute  $b:=[[2^{k^*(v(\Sc))}v(\Sc)]_M]_2$.
\item[Step 7.]\label{a8}
  If $b=0$ then output `yes' else output `no'.
  \end{description}

\textbf{Claim 1.} The algorithm always gives a correct answer.

\begin{proof}
  Suppose that the algorithm finishes at Step~5. 
  If $v(\Sc)>0$ then $v(\Sc)=[v(\Sc)]_M$ which implies
  $[v(\Sc)]_2=[[v(\Sc)]_M]_2$. If $v(\Sc)<0$ then
  $[v(\Sc)]_2+[[v(\Sc)]_M]_2=[M]_2=1$. In both cases the algorithm
  gives a correct answer.

  Suppose that the algorithm finishes at Step~7. Note that in this
  case  $k^*(v(\Sc))>0$. So, Lemma~\ref{criterion} can be applied. 
  The algorithm gives a correct answer due to the following observation:
  $x>0$  iff  $[x]_M<M/2$ assuming  $0\leq x <M/2$.
\end{proof}

\textbf{Claim 2.} The algorithm runs in polynomial memory.

\begin{proof}
  Computing $[v(\Sc)]_2$ takes a polynomial time as mentioned above.
  The algorithm computes $[[2^kv(\Sc)]_M]_2$ by using
  Eq.~\eqref{Chinese} modulo~2
  \begin{equation}\label{Chinese2}
    \sum_{i=1}^m \xi_i \equiv\rho(x)+[x]_M \pmod2.
  \end{equation}
  To compute $[x]_M\bmod2$ we need to compute $\xi_i$ and
  $\rho(x)$. Computing $2^kv(\Sc)\bmod n$ is possible in polynomial
  time. To compute $M_i^{-1}\bmod p_i$ one can take all primes from
  $3$ to $2^{2\ell}$ one by one, compute an inverse residue
  modulo~$p_i$ and multiply it by the current product. 
  This  as well as summation of $\xi_i$ can be done in polynomial memory.

  For each $i$ an
  $(2^{-2\ell})$-approximation to $\xi_i/p_i$ is computed in polynomial
  time. Summation of these approximations is directly implemented in
  polynomial memory. So, it is possible to compute $\lfloor
  \sigma(2^{k}v(\Sc))\rfloor$ and 
  $\{\sigma(2^{k}v(\Sc))\}$ in polynomial memory (for
  $k=2^{\poly(\ell)}$). Starting 
  from $k=0$ and incrementing $k$ until
  $k^*(v(\Sc))$ is found may be implemented in polynomial
  memory. Having all these data, it is easy to compute 
  $\rho(2^{k^*(v(\Sc))}v(\Sc))$ and  $[[2^{k^*(v(\Sc))}v(\Sc)]_M]_2$.
\end{proof}

\section{Open questions}\label{open}

The main open question here is the complexity of SDFP. 
We suggest the problem $\AC_\geq$ as a `lowerbound' for SDFP. If it is
hard then SDFP is hard too. The maximal result in this direction would
be \PSPACE-hardness of SDFP. 
Thus from the complexity viewpoint it is very interesting to put
SDFP in \PSPACE if possible. 

Complexity of $\AC_\geq$ itself is the next question. To be a good
`lowerbound' it should be hard. But up to our current knowledge
nothing prohibits inventing an 
efficient algorithm for its solution.
Such an algorithm would be interesting by other reasons. It would
justify using circuit representation for rationals instead of binary
system in complexity theoretic questions when time is estimated up to
a polynomial slowdown. It might simplify complexity analysis of
numeric algorithms very much. So, any definite result about $\AC_\geq$
would lead to interesting conclusions.

From algorithmic point of view some intrinsic difficulty of numerical
algorithms is related to nonconstructive nature of real
numbers. Though some computational models, e.g. BSS model of
computation over arbitrary field initially proposed by
Smale~\cite{BCSS}, directly operate with real numbers and thus
completely ignore the question of number representation.  Presently
the relation of BSS model to the common model of algorithmic
complexity is not well understood.  In particular, in BSS model SDFP
falls into the complexity class $\NP_\RR$ --- an analogue of the class
\NP.  Reffering to our way of number representation it is natural to
ask about the inclusion $\mathrm{SDFP}\in\NP^{\AC_\geq}$. This
question is also open.

In the opposite direction, it may be interesting to represent numbers
as optimal solutions of SDP. For example, the well known $n!$
conjecture by Shub and Smale~\cite{SS} in our setting asks about
monotone circuit complexity of $n!$. It follows from the reduction in
Theorem~\ref{upper} that monotone circuit complexity is lowerbounded
by the dimension of a SDP representing $n!$ as an optimal solution. Is
this dimension polylogarithmic on $n$?

\section*{Acknowledgements}

We would like to thank A. Kitaev for helpful discussions.
The work of both authors was  supported by RFBR
grant 05--01--01019. The second author was also supported by grant 
NSh 1721.2003.1. 

And the last but not the least: we are much indebted to our late friend
and teacher L.Khachiyan who stimulated our interest to this problem.

\end{document}